\documentclass[3p,times]{elsarticle}

\usepackage{ecrc}

\volume{00}

\firstpage{1}

\journalname{Nuclear Physics A}
\runauth{E. Iancu}

\jid{npa}

\jnltitlelogo{Nuclear Physics A}

\usepackage{amssymb}
\usepackage[figuresright]{rotating}
\usepackage{bm,amsmath,amssymb}

\biboptions{square,comma,numbers,sort&compress}

\usepackage[figuresright]{rotating}

\long\def\comment#1{ }

\def\lesssim{\mathrel{
   \rlap{\raise 0.511ex \hbox{$<$}}{\lower 0.511ex \hbox{$\sim$}}}}
   \def\gtrsim{\mathrel{%
   \rlap{\raise 0.511ex \hbox{$>$}}{\lower 0.511ex \hbox{$\sim$}}}}

\begin{document}

\begin{frontmatter}

\dochead{}

\title{Di--jet asymmetry and wave turbulence}

\author{Edmond Iancu}
\ead{edmond.iancu@cea.fr}

\address{Institut de Physique Th\'{e}orique de Saclay, F-91191 Gif-sur-Yvette, France}

\begin{abstract}
We describe a new physical picture for the fragmentation of an energetic jet propagating
through a dense QCD medium, which emerges from perturbative QCD and has the potential
to explain the di--jet asymmetry observed in Pb--Pb collisions at the LHC. 
The central ingredient
in this picture is the phenomenon of wave turbulence, which provides a very efficient mechanism
for the transport of energy towards the medium, via many soft particles which
propagate at large angles with respect to the jet axis.
\end{abstract}

\begin{keyword}
Perturbative QCD, Heavy Ion Collisions, Jet Quenching, Wave Turbulence
\end{keyword}

\end{frontmatter}


\section{Introduction}
\label{}

One of the most interesting discoveries of the heavy ion program at the LHC is the
phenomenon known as {\em di--jet asymmetry} --- a strong imbalance between the energies of 
two energetic back--to--back jets produced in an ultrarelativistic nucleus-nucleus collision.
This is attributed to the effect of the interactions of one of the two jets with the 
dense QCD matter that it traverses, while the other leaves the system unaffected. 
Originally identified \cite{Aad:2010bu}
as missing transverse energy within a conventionally defined `jet' with small angular opening 
(the same as for the trigger jet),  this phenomenon has been subsequently shown,
via more detailed studies \cite{Chatrchyan:2011sx,Aad:2012vca}, to consist in the transport of a 
part of the jet energy towards large angles and by soft particles.
The total amount of energy thus transferred from small to large angles 
(about 10 to 20~GeV) is considerably larger than the typical transverse 
momentum, $\sim 1$~GeV, of a parton in  the 
medium, so in that sense the effect is large and potentially non--perturbative.

Yet, there exists a mechanism within perturbative QCD which 
naturally leads to energy loss at large angles: the BDMPSZ  mechanism for 
medium--induced gluon radiation \cite{Baier:1996kr,Zakharov:1996fv}. 
Most previous studies within this approach 
have focused on the energy lost by the leading particle, as controlled by relatively hard
emissions at small angles. More recently, in the wake
of the LHC data, the attention has been shifted towards softer emissions (which occur at large angles) 
and, more generally, towards a global understanding of the in--medium
jet evolution.
This raised the difficulty of including the effects of multiple gluon branchings, which become
important for the soft emissions.
After first studies of interference phenomena, which exhibited the role of
medium rescattering in destroying the colour coherence between partonic sources
\cite{MehtarTani:2010ma,CasalderreySolana:2011rz},
we have recently demonstrated \cite{Blaizot:2012fh} 
that the in--medium jet evolution can be reformulated (to the perturbative accuracy of interest) as
a classical stochastic process. This allows for systematic numerical studies  
via Monte Carlo methods, like for jets fragmenting in the vacuum. It also allows
for analytic studies, at least for particular problems, like the recent study of the energy flow
throughout the cascade in Ref.~\cite{Blaizot:2013hx}. This study revealed a remarkable
phenomenon, which is new in the context of QCD and which, besides its conceptual interest, 
has also the potential to explain the LHC data for di-jet asymmetry: the  {\em wave turbulence}. 
The developments in Refs.~\cite{Blaizot:2012fh,Blaizot:2013hx} 
 will be briefly reviewed in what follows, with emphasis on the physical picture of wave turbulence.

\section{Medium--induced radiation \`a la BDMPSZ}

The BDMPSZ mechanism relates the radiative energy loss
by an energetic parton propagating through a dense QCD medium (`quark--gluon plasma') to the
transverse momentum broadening via scattering off the medium constituents. A central
concept is the {\em formation time} $\tau_{_{\rm br}}(\omega)$ --- the typical times it takes 
a gluon with energy $\omega \ll E$ to be emitted.
($E$ is the energy of the original parton, a.k.a. the `leading particle'.) The gluon starts as a virtual 
fluctuation which moves away from its parent parton via quantum diffusion:
the transverse\footnote{The `transverse directions' refer to the 2--dimensional plane orthogonal 
to the 3--momentum of the leading particle (the `longitudinal axis').}
separation $b_\perp$ grows with time as $b_\perp^2 \sim \Delta t/\omega$.
The gluon can be considered as `formed' when it loses coherence w.r.t to its source, meaning that  
$b_\perp$ is at least as large as the gluon transverse wavelength $\lambda_\perp
= 1/k_\perp$.  But the gluon transverse momentum $k_\perp$ is itself increasing with time, 
via collisions  which add random kicks $\Delta k_\perp$
at a rate given by the {\em jet quenching parameter} $\hat q$ :
$\Delta k_\perp^2\sim \hat q \Delta t$. The `formation' condition, $b_\perp\gtrsim 1/\Delta k_\perp$ for
$\Delta t\gtrsim\tau_{_{\rm br}}$, implies
\begin{eqnarray}\label{thetaf} 
\tau_{_{\rm br}}(\omega)\,\simeq \,\sqrt{\frac{2\omega}{\hat q}}\,, \qquad  k_{_{\rm br}}^2 \,=\,\hat q \tau_{_{\rm br}}(\omega)
\,\simeq\,
 (2\omega\hat q)^{1/2}\,,\qquad \theta_{_{\rm br}}\,\simeq\,\frac{k_{_{\rm br}}}{\omega}\,\simeq\,
 \left(\frac{2 \hat q}{\omega^3}\right)^{1/4},\end{eqnarray}
where $k_{_{\rm br}}$ and $\theta_{_{\rm br}}$ are the typical values of the
gluon transverse momentum and its emission angle at the time of formation.
Eq.~(\ref{thetaf}) applies as long as $\ell\ll \tau_{_{\rm br}}(\omega) < L$, where $L$ is the length of the medium
and $\ell$ is the mean free path between successive collisions.
The second inequality implies an upper limit on the energy 
of a gluon that can be emitted via this mechanism, and hence a lower limit on
the emission angle: $\omega \lesssim \omega_c\equiv \hat q L^2/2$ and 
$\theta_{_{\rm br}}\gtrsim  \theta_c\equiv 2/(\hat q L^3)^{1/2}$. The BDMPSZ regime
corresponds to $\hat q L^3\gg 1$ and hence $\theta_c\ll 1$. Choosing $\hat q=1\,{\rm GeV}^2/{\rm fm}$ 
(the weak coupling estimate \cite{Baier:1996kr}  for a QGP with temperature 
$T=250$~MeV) and $L= 4$~fm, one finds
$ \omega_c\simeq 40$~GeV and $\theta_c\simeq 0.05$.

It is furthermore easy to deduce a parametric estimate for the spectrum of the emitted gluons
(at least for the relatively soft gluons with $\omega\ll \omega_c$): this is the product of the standard
bremsstrahlung spectrum for the emission of a single gluon times the average number of emissions
which can occur within the plasma, that is 
 $L/\tau_{_{\rm br}}$~:
  \begin{eqnarray}\label{spec}
\omega \frac{{\rm d} N}{{\rm d} \omega}
 \,\simeq\,\frac{\alpha_s N_c}{\pi}\,\frac{L}{\tau_{_{\rm br}}(\omega)}\,=\,
 \bar\alpha\sqrt{\frac{\omega_c}{\omega}}\,, \end{eqnarray}
with $\bar\alpha\equiv \alpha_s N_c/\pi$. 
Note that the number of emissions $L/\tau_{_{\rm br}}$
is much smaller than the number of collisions $L/\ell$, since several successive collisions can coherently
contribute to a single emission; this is known as the LPM  effect
(Landau, Pomeranchuk, Migdal) and leads to the characteristic $\sim 1/\sqrt{\omega}$
dependence of the BDMPSZ spectrum \eqref{spec}. By integrating this spectrum over all the energies
$\omega\le\omega_c$, one estimates to the total energy loss by the leading particle:
\begin{eqnarray}\label{Etot}
\Delta E_{\rm tot}=\int^{\omega_c} {\rm d} \omega\
  \omega\,\frac{{\rm d} N}{{\rm d} \omega}\,
 \sim\, \bar{\alpha}\omega_c\,\sim\, \bar{\alpha}\hat q L^2\,.\end{eqnarray}
The above integral is dominated by its upper limit: the total energy loss is controlled
by the hardest possible emissions, those with energies $\omega\sim \omega_c$. Such hard emissions,
however, propagate at small angles $\theta\sim \theta_c$ w.r.t. to the jet axis, so they remain
a part of the conventionally defined `jet' and thus cannot contribute to the di-jet asymmetry.
On the other hand, the soft gluons with $\omega\ll \omega_c$ are emitted directly at large angles
$\theta\gg \theta_c$ and, moreover, these angles are further enhanced after emission via rescattering
in the medium:  a gluon which crosses the medium over a distance $\sim L$ acquires a transverse 
momentum broadening $k_\perp^2 \sim \hat q L\equiv Q_s^2$, which for $\omega\ll \omega_c$ 
is in fact  larger than the respective momentum acquired during formation: $Q_s^2\gg
k_{_{\rm br}}^2(\omega)$. Accordingly, a soft 
gluon emerges at a typical angle $\theta(\omega)\sim Q_s/\omega$ which is even larger
than $\theta_{_{\rm br}}(\omega)$ --- and of course much larger than $\theta_c$. It is interesting
to try and estimate the typical energy which would be transported in this way 
at angles larger than a given value $\theta_0$, with  $\theta_0\gg \theta_c$~:
 \begin{eqnarray}\label{Etheta0}
\Delta E(\theta > \theta_0)=\int^{\omega_0} {\rm d} \omega\
  \omega\,\frac{{\rm d} N}{{\rm d} \omega}\,
 \sim\, \bar{\alpha}\sqrt{\omega_c\omega_0}\,\propto  \,\frac{1}{\sqrt{\theta_0}}\quad\mbox{with}\quad 
  \omega_0\,\equiv\,\frac{Q_s}{\theta_0}\,.
  \end{eqnarray}
This is only a small fraction $(\theta_c/\theta_0)^{1/2}$ of the total energy loss \eqref{Etot},
but it is lost at large angles, so it counts for the energy loss {\em by the jet}. Yet, Eq.~(\ref{Etheta0})
does not show the right trend to explain the LHC data: this estimate decreases quite fast with
increasing $\theta_0$, thus predicting that most of the energy loss should lie {\em just outside} the jet cone
(and thus be easily recovered when gradually increasing the jet angular opening).
This contradicts the results of a detailed analysis by CMS  \cite{Chatrchyan:2011sx}, which show that  
most of the 
`missing' energy is deposited at very large angles $\theta > 0.8$.

However, the previous argument misses an important ingredient : the gluon spectrum
\eqref{spec} is a measure of the probability for emitting a gluon via the BDMPSZ mechanism.
For $\omega\sim \omega_c$, this probability if of $\mathcal{O}(\bar{\alpha})$, showing that hard emissions are
relatively rare events. But when $\omega\sim\bar{\alpha}^2 \omega_c$, this probability becomes
of $\mathcal{O}({1})$, meaning that the soft emissions with $\omega\lesssim \bar{\alpha}^2\omega_c$ can occur
abundantly, event-by-event. For such small energies, the result \eqref{spec} must be corrected 
to account for  {\em multiple emissions} and, especially, {\em multiple branchings} of the soft emitted gluons
\cite{Blaizot:2012fh,Blaizot:2013hx}.

\section{Democratic branchings and wave turbulence}

\begin{figure}[t]
	\centerline{\includegraphics[width=0.5\linewidth]{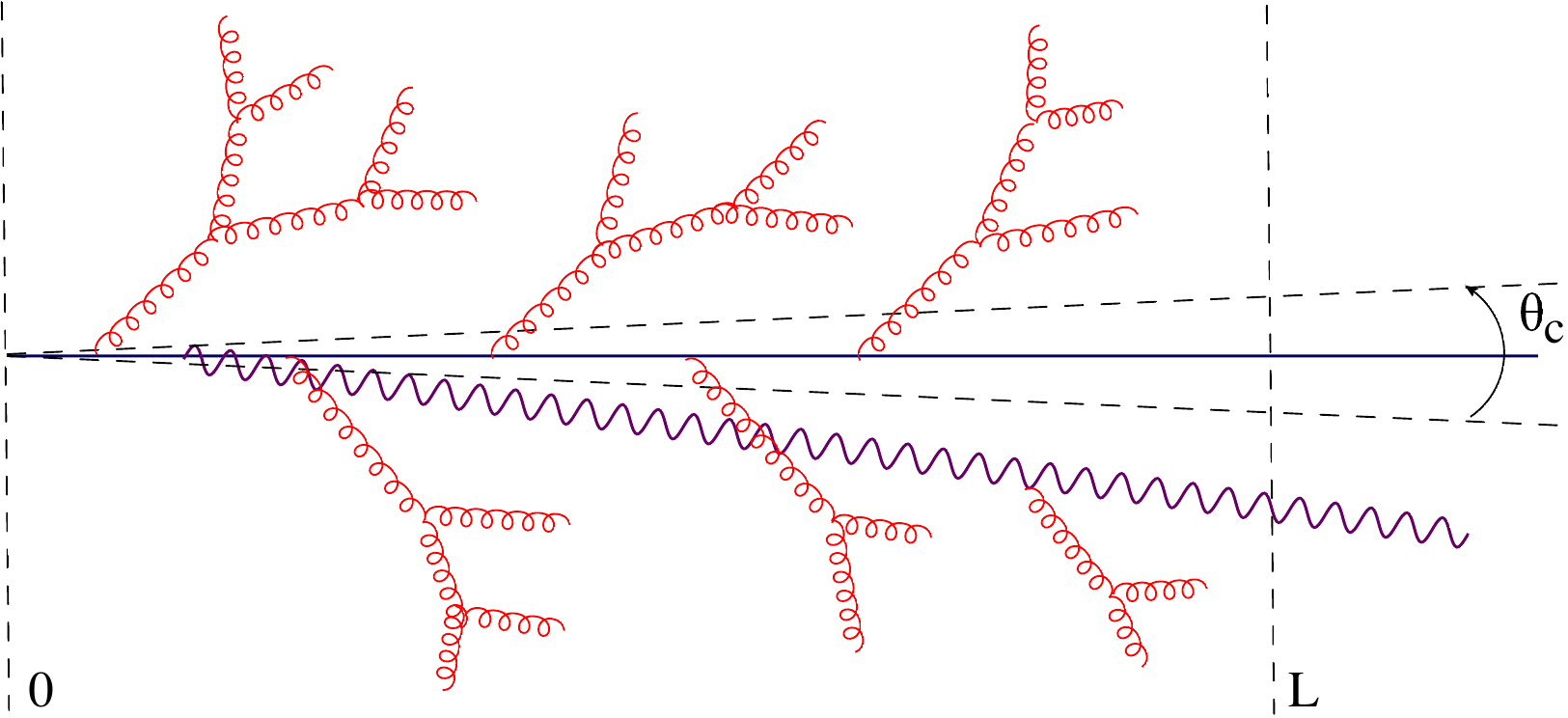}}
	\caption{A medium--induced cascade: the  leading particle emits one 
	hard gluon with $\omega \sim\omega_c$ (at
a small angle $\theta\sim \theta_c$) together with a myriad of soft gluons
with $\omega \lesssim\alpha_s^2 \omega_c$, which in turn generate gluon
cascades (at relatively large angles) via successive, quasi--democratic, branchings.}
\label{fig:jet}
\end{figure}

Multiple soft emissions by the leading particle have already been discussed by BDMPS \cite{Baier:2001yt}:
they change the energy distribution of the leading hadron, but not the inclusive spectrum
\eqref{spec} for the soft radiation, nor its (unrealistic) prediction for the angular distribution of the
radiation, Eq.~(\ref{Etheta0}). What is more important for the present purposes, is the fate of the soft
gluons after being emitted. The probability for a gluon with energy $\omega$ to split into two daughter
gluons with energy fractions $x$ and $1-x$ is obtained by replacing $\omega\to x(1-x)\omega$
in Eq.~(\ref{spec}). Hence, when  $\omega\sim\bar{\alpha}^2 \omega_c$, this probability is of $\mathcal{O}({1})$ for
{\em generic} values of $x$ : the soft parent gluon is {\em certain} to split and its
branching is {\em quasi--democratic} (i.e. unbiased towards the endpoints at $x=0$
and $x=1$ of the distribution in $x$) \cite{Blaizot:2013hx}. 
For even softer energies, $\omega \ll \bar{\alpha}^2\omega_c$, the
{\em lifetime} $\Delta\tau$ of a gluon generation, i.e.
the time interval  between two successive branchings, is considerably smaller than the medium 
size\footnote{This estimate for $\Delta\tau$ follows from the condition that
the emission probability
$\mathcal{P}(\omega)\simeq \bar{\alpha}(\Delta\tau/\tau_{_{\rm br}}(\omega))$ become of $\mathcal{O}({1})$.}: 
$\Delta\tau\sim (1/\bar{\alpha})\tau_{_{\rm br}}(\omega)\ll L$.
Hence, such soft gluons undergo successive branchings leading to gluon cascades. 
Being quasi--democratic, these branchings efficiently degrade the
energy to smaller and smaller values of $x$. And since the gluons produced by these branchings
are softer and softer, they get easily deviated by the collisions in the medium to larger and larger angles
(see Fig.~\ref{fig:jet}).
Thus, the quasi--democratic and quasi--deterministic cascade provides a very efficient mechanism
for transporting energy at large angles. This mechanism is a manifestation of a phenomenon 
well known in other fields of physics : the {\em wave turbulence} \cite{KST,Nazarenko}.

Before we characterize this new phenomenon in mode detail, let us describe the formalism which allows
us to treat multiple branching \cite{Blaizot:2012fh}. 
In principle, one can construct a parton cascade by iterating the $1\to 2$
`vertex' for parton splitting, which here is the BDMPSZ spectrum \eqref{spec}.
This would certainly be the correct procedure for a {\em classical} branching process. It turns out to
that this is also the right procedure for the  {\em quantum} problem at hand, but in this case
such a procedure is highly non-trivial, as it could be invalidated by {\em interference
phenomena}. Recall e.g. the evolution of a jet via successive parton branching {\em  in the vacuum}: 
the daughter
partons produced by one splitting remain `color-coherent' with each other (their
total color charge is fixed to be equal to the respective charge of the parent parton) until the next 
splitting of any of them. This coherence implies interferece effects between
the emissions by the two daughter partons, which in that context
are well known to be important: they lead to the  {\em angular ordering} of successive
emissions, which ultimately favors jet collimation \cite{Dokshitzer:1991wu}.

Remarkably, the situation in that respect appears to be simpler for 
parton branching {\em in the medium} \cite{MehtarTani:2010ma,CasalderreySolana:2011rz,Blaizot:2012fh}:
the daughter partons efficiently randomize their color charges via rescattering
in the medium and thus lose their mutual  
color coherence already during the formation process \cite{Blaizot:2012fh}.
Accordingly, the interference effects are suppressed (as compared to the independent branchings)
by a phase--space factor  $\tau_{_{\rm br}}(\omega)/L$, which is small whenever $\omega\ll \omega_c$.
This implies that the successive medium--induced emissions can be effectively
treated as {\em independent} of each other and taken into account via a {\em probabilistic 
branching process},  in which the BDMPSZ spectrum plays the role of a
branching rate. 
Such a process has already been used in applications to phenomenology, albeit on
a heuristic basis \cite{Baier:2000sb,Baier:2001yt,Jeon:2003gi}.

The general branching process is a Markovian process in 3+1 dimensions which describes the
gluon distribution in energy ($\omega$) and transverse momentum ($k_\perp$), and
its evolution when increasing the medium size $L$ (see Ref.~\cite{Blaizot:2012fh} for details). 
This process is well
suited for numerical studies via Monte-Carlo simulations. But analytic results have been
obtained too \cite{Blaizot:2013hx}, for a simplified process in 1+1 dimensions
which describes the energy distribution alone. 
These results lead to an interesting physical picture, that of wave turbulence, that we shall now describe.

\begin{figure}[t]
	\centerline{\includegraphics[width=0.7\linewidth]{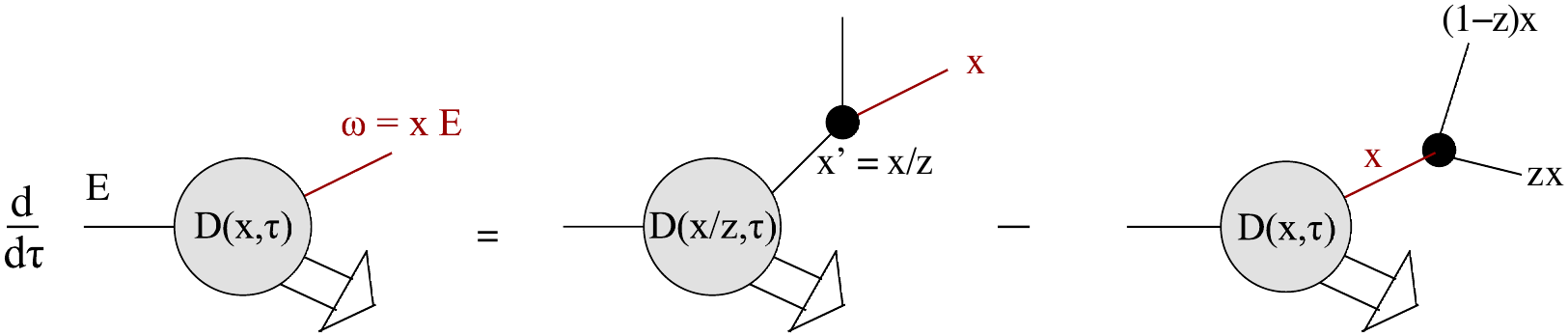}}
	\caption{The change in  the gluon spectrum $D(x,\tau)\equiv x({{\rm d} N}/{{\rm d} x})$
	due to one additional branching $g\to gg$.}
\label{fig:rate}
\end{figure}

To that aim, it is convenient to focus on the {\em gluon spectrum} 
$D(x,\tau)\equiv x({{\rm d} N}/{{\rm d} x})$, where $x\equiv \omega/E$ is the energy fraction carried by 
a gluon from the jet and the `evolution time' $\tau$ is the medium size in 
dimensionless units, as defined in the equation below. 
The quantity $D(x,\tau)$ obeys a `rate' equation \cite{Blaizot:2013hx,Baier:2000sb,Jeon:2003gi},
which reads, schematically,
\begin{eqnarray}\label{rate}
\frac{\partial D(x,\tau)}{\partial\tau}\,= \,{\cal I}[D](x,\tau)\,\equiv\,\mathrm{Gain}[D]-\mathrm{Loss}[D]
\,,\qquad{\rm with}\quad
 \tau\,\equiv\,\bar \alpha\sqrt{\frac
 {\hat q}{E}}\,L\,.
\end{eqnarray}
where the `collision term' ${\cal I}[D]$ (a linear functional of  $D(x,\tau)$) is the difference between
a `gain' term and a `loss' term, as illustrated in Fig.~\ref{fig:rate}.
The `gain' term describes the increase in the number of gluons with a given $x$
via radiation from gluons with a larger $x'=x/z$, with any $x<z<1$. The `loss' term expresses the 
decrease in the number of gluons at $x$ via their decay $x\to zx, (1-z)x$, with any $0<z<1$.
By construction, the first iteration of this equation coincides with the BDMPSZ spectrum \eqref{spec}, 
which in our new notations reads (for relatively soft gluons with $x\ll 1$)
 \begin{eqnarray}\label{Ds}
 D^{(1)}(x\ll 1,\tau)\,\simeq\,\frac{\tau}{\sqrt{x}}\,. \end{eqnarray}
This approximation breaks down when $D^{(1)}(x,\tau)\sim\mathcal{O}({1})$, meaning for $x\lesssim \tau^2$
(the familiar condition $\omega \lesssim \bar{\alpha}^2\omega_c$ in these new notations).
In this non--perturbative regime at small $x$, one needs an exact result which resums 
multiple branchings to all orders. Such a solution has been presented in \cite{Blaizot:2013hx} and reads
(for $x\ll 1$ once again)
 \begin{eqnarray}\label{Dex}
 D(x\ll 1,\tau)\,\simeq\,\frac{\tau}{\sqrt{x}}\ {\rm e}^{-\pi{\tau^2}}\,.\end{eqnarray}
Formally, one can read Eq.~(\ref{Dex}) as `BDMPSZ spectrum by the leading particle $\times$ survival
probability for the latter'. However, unlike Eq.~(\ref{Ds}), the spectrum (\ref{Dex}) also includes the effects of
multiple branchings. That is, the energy in a given bin with $x\ll 1$ is produced both via
direct radiation by the leading particle, and via energy transfer from the higher
bins at $x'>x$, through successive splittings. The persistence of the scaling spectrum $D_s\equiv
1/\sqrt{x}$ under this evolution demonstrates that this spectrum  is a {\em fixed point} of the collision kernel:  
${\cal I}[D_s](x)=0$ for $x\ll 1$. In turn, this means that the rate for energy transfer from one parton 
generation to the next one is independent of the generation (i.e. of $x$). This property
is the distinguished signature of {\em wave turbulence}  \cite{KST,Nazarenko}:
 via successive splittings,
the energy {\em flows} from large $x$ to small $x$ without accumulating at any intermediate value of $x$.
It rather accumulates into a condensate at $x=0$. Since there is only a finite amount of energy
available (the energy $E$ of the leading particle), it follows 
that the total energy which is contained in the spectrum (i.e. in the bins at $0<x\le 1$) 
must decrease with time. Indeed, a direct calculation yields 
 \cite{Blaizot:2013hx}
\begin{eqnarray}\label{Eflow} \int_0^1 {\rm d} x \,D(x,\tau)\,=\,{\rm e}^{-\pi\tau^2}\quad \Longrightarrow\quad
  {\cal E}_{\rm flow}(\tau)\,\equiv \,1-\int_0^1 {\rm d} x D(x,\tau)\,=\, 1- {\rm e}^{-\pi\tau^2}\,.\end{eqnarray}
The quantity ${\cal E}_{\rm flow}(\tau)$  is the energy fraction carried away
by the flow and which formally ends up in the condensate. As we shall shortly discuss,
this energy is in fact transferred to the medium, at very large angles.

\begin{figure}[t]
	\centering
	\includegraphics[width=0.5\linewidth]{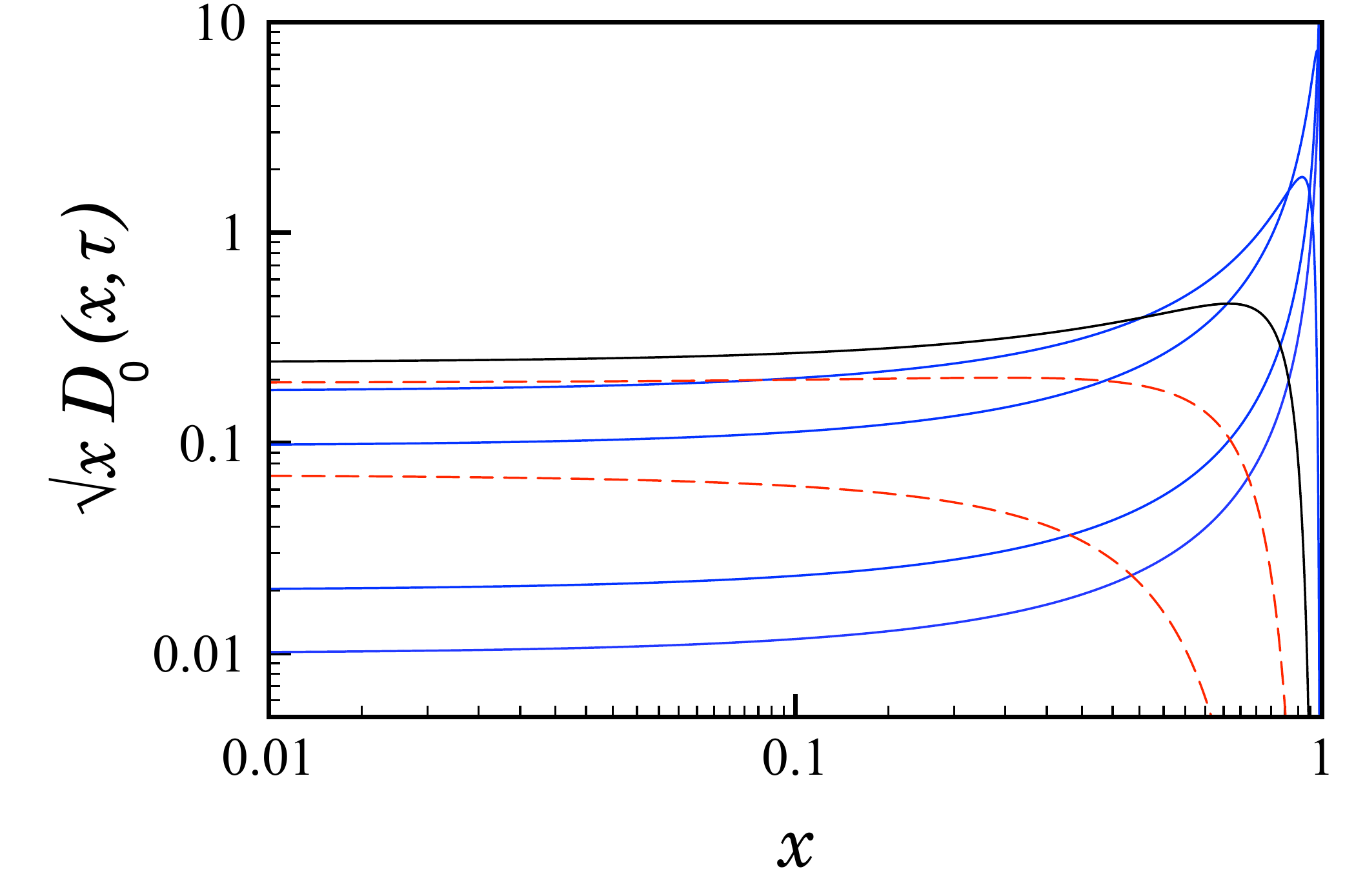}
		\caption{Plot (in Log-Log scale) of $\sqrt{x} D(x,\tau)$ as a function of $x$ for various values of $\tau$ 
		(full lines from bottom to top: $\tau=0.01,0.02,0.1,0.2,0.4$; dashed lines from top down: $\tau=0.6,0.9$);
		from Ref.~\cite{Blaizot:2013hx}.
		}
		\label{fig:spec}

\end{figure}

These considerations are
illustrated in Fig.~\ref{fig:spec} which shows the spectrum for various values of $\tau$.
At small $\tau\ll 1/\sqrt{\pi}$, the small--$x$ part of the spectrum rises linearly with $\tau$, 
as shown by Eq.~(\ref{Ds}) (see the full lines in Fig.~\ref{fig:spec}).
At the same time, the leading--particle peak, which originally was a $\delta$--function
at $x=1$, moves at $1-x\simeq \pi\tau^2$ and becomes
broader. For larger times $\tau\gtrsim 1/\sqrt{\pi}$, the source disappears 
and the spectrum is globally suppressed
by the Gaussian factor in (\ref{Dex}); yet, the scaling behavior $D\propto 1/\sqrt{x}$ is still visible
at small $x$ (see the dotted lines in Fig.~\ref{fig:spec}).

\section{Energy loss at large angles}

The emergence of a flow component  ${\cal E}_{\rm flow}(\tau)$ in the energy transport
down the cascade explains one of the main characteristics of (wave) turbulence: this is a very efficient 
mechanism for transferring energy  between two widely separated scales --- here, from $x=1$
down to $x=0$. To see this, let us compute the energy transferred after time $\tau$ 
below a given value $x_0\ll 1$. This includes two components: the energy which is contained in
the spectrum, in the bins at $0<x<x_0$, and the flow energy, which is independent of $x_0$
(since accumulated at $x=0$). Thus,
\begin{eqnarray}\label{Esmall}
  {\cal E}(x \le x_0,\tau) \,=\,2\tau \sqrt{x_0}\,{\rm e}^{-\pi\tau^2}\,+\,
  (1- {\rm e}^{-\pi\tau^2})\,\simeq\,2\tau \sqrt{x_0}\,+\,\pi\tau^2\,,
    \end{eqnarray}
where the second, approximate, equality holds for $\pi\tau^2\ll 1$. Note that,
even for small times, the flow component dominates over the
non--flow one provided $x_0 < \tau^2$, that is, in the non--perturbative regime at small $x$
where the multiple branching becomes important. For larger times $\tau\gtrsim 1/\sqrt{\pi}$,
the flow piece dominates for any $x_0$ and approaches unity, meaning that the whole energy
can be lost towards arbitrarily soft quanta, which propagate at arbitrarily large angles.

To understand how remarkable this situation is, let us compare it with the more familiar
example of the DGLAP evolution (say, for a jet in the vacuum), where there is {\em no flow}. 
(The DGLAP equation too can be viewed
as a `rate equation',  cf. Eq.~(\ref{rate}), with the logarithm of the virtuality playing the
role of the `evolution time'.)
In that case, the splittings are typically asymmetric 
($x\to 0$ or $x\to 1$), leading to a rapid increase in the number of gluons at small $x$. Yet
most of the energy remains in the few partons with relatively large values of $x$. Indeed,  for the
DGLAP cascade, the energy is fully contained within the spectrum (no flow) 
and the energy sum-rule  $\int_0^1 {\rm d} x D(x,\tau)=1$
is dominated by the higher values of $x$ in the support of the function $D(x,\tau)$ at time $\tau$.
Conversely,
one can show that a necessary condition for the emergence of (turbulent) flow is quasi--democratic
branching \cite{KST}.

So far, we have assumed that the evolution remains unchanged down to $x=0$, but physically
this is not the case: when the gluon energies become
as low as the typical energy scale in the medium --- the `temperature' $T\sim 1$~GeV ---, 
the gluons  `thermalize' and disappear from the jet. The energy which is thus
transferred to the medium (and hence lost by the jet) can be evaluated by replacing $x_0\to x_{\rm th}
\equiv T/E$ in Eq.~(\ref{Esmall}). This energy loss is independent of the details of the thermalization  
mechanism and even of the medium temperature (since dominated by the flow component,
as we shall shortly see). This {\em universality} is the hallmark of turbulence:
the rate for energy transfer at the lower end of the cascade is fixed by the turbulent flow
alone, and thus is independent of the specific mechanism for dissipation. 

To make contact with the phenomenology, we notice that for a jet with
$E=100~{\rm GeV}\approx 2\omega_c$, Eq.~(\ref{rate}) implies $\tau\equiv\bar{\alpha}\sqrt{2\omega_c/E}
\simeq\bar{\alpha}\simeq 0.3$, which is quite small.
The flow piece in Eq.~(\ref{Esmall}), which is independent of $x_0$, dominates over the
non--flow piece for any $x_0 < \tau^2\simeq 0.1$, a value much larger than the thermalization 
scale $x_{\rm th}\simeq 0.01$. Thus, in evaluating the energy loss via thermalization, one can keep
only the flow component in the small--$\tau$ version of Eq.~(\ref{Esmall}), as anticipated. 
Returning to physical units, one finds
  \begin{eqnarray}\label{Eth}
\Delta E_{\rm th}\,\simeq\,E\,{\cal E}_{\rm flow}\,\simeq\,{\upsilon}\,\bar{\alpha}^2 \omega_c
 \,,\end{eqnarray}
where $\upsilon$ would be equal to $2\pi$ according to Eq.~(\ref{Esmall}), but a more precise
calculation yields $\upsilon\simeq 4.96$  \cite{Blaizot:2013hx}. This is formally suppressed by an
additional power of $\bar{\alpha}$ as compared to the total energy loss by the leading particle, Eq.~(\ref{Etot}),
which we recall is controlled by hard gluon emissions ($\omega\sim \omega_c$) at small angles
($\theta\sim\theta_c$). However, the flow contribution in Eq.~(\ref{Eth}) is numerically quite large
(because $\upsilon$ is a reasonably large number) and moreover this is associated
with soft emissions at large angles. It thus has the potential to explain the LHC data for
di--jet asymmetry. 

With $\omega_c=40$~GeV, Eq.~(\ref{Eth}) predicts $\Delta E_{\rm th}\simeq 20$~GeV, a value that 
compares well with the experimental observations. This energy is carried by the relatively
soft quanta at the lower end of the cascade ($x\sim x_{\rm th}$), that is, by particles whose
energies are comparable to the `temperature' $T$ of the medium. Precisely because they are
so soft, these particles propagate at very large angles with respect to the jet axis.
To obtain a parametric estimate for these angles, we recall that a
gluon with energy $\omega \lesssim \bar{\alpha}^2 \omega_c$ has
a lifetime $\Delta\tau\sim (1/\bar{\alpha})\tau_{_{\rm br}}(\omega)\lesssim L$, during which
it accumulates a transverse momentum broadening $k_\perp^2 \sim \hat q \Delta \tau = (1/\bar{\alpha})
k_{_{\rm br}}^2$, via collisions in the medium. Accordingly, this gluon should emerge 
at an angle (compare to Eq.~(\ref{thetaf}))
\begin{eqnarray}
\theta(\omega)\, \sim \, \frac{1}{\sqrt{\bar{\alpha}}}\,\theta_{_{\rm br}}(\omega)\,\sim\,
 \frac{1}{\sqrt{\bar{\alpha}}}\left(\frac{2 \hat q}{\omega^3}\right)^{1/4}\,.
 \end{eqnarray}
A {\em lower limit} on this angle is obtained by choosing
$\omega\sim \bar{\alpha}^2 \omega_c\sim 4$~GeV (the non--perturbative energy scale 
below which develops the turbulent cascade); this yields 
$\theta\simeq 0.5$. But for a  {\em typical}  gluon with
 $\omega\sim T\sim 1\div 2$~GeV, this angle is even larger: $\theta\sim \mathcal{O}({1})$.
 This is in qualitative and even quantitative agreement with the detailed analyses of the data by
CMS \cite{Chatrchyan:2011sx} and ATLAS \cite{Aad:2012vca}, which show that most
of the  `missing' energy lies at very large angles $\theta\gtrsim 0.8$.

{\bf Acknowledgements}
This research is supported by the European Research Council under the Advanced Investigator Grant ERC-AD-267258.

\end{document}